\documentclass[useAMS,usenatbib]{mn2e}

\usepackage{graphicx}
\usepackage{color}

\title[The Host of GRB\,080517]{GRB 080517: A local, low luminosity GRB in a dusty galaxy at z=0.09}
\author[E.~R.~Stanway et al.]{Elizabeth R. Stanway$^{1}$\thanks{E-mail:
e.r.stanway@warwick.ac.uk}, Andrew J. Levan$^{1}$, Nial Tanvir$^{2}$, Klaas Wiersema$^{2}$, \newauthor  Alexander van der Horst$^3$, Carole G. Mundell$^4$, Cristiano Guidorzi$^5$\\
$^{1}$Department of Physics, University of Warwick, Gibbet Hill Road, Coventry, CV4 7AL, UK\\
$^{2}$Department of Physics and Astronomy, University of Leicester, University Road, Leicester LE1 7RH, UK\\
$^{3}$Anton Pannekoek Institute, University of Amsterdam, Science Park 904, 1098 XH Amsterdam, The Netherlands\\
$^{4}$Astrophysics Research Institute, Liverpool John Moores University, IC2, Liverpool Science Park, 146 Brownlow Hill, Liverpool L3 5RF, UK\\
$^{5}$Department of Physics and Earth Sciences, University of Ferrara, via Saragat 1, I-44122, Ferrara, Italy\\}

\begin{document}

\date{Accepted 2014 October 28.  Received 2014 October 23; in original form 2014 September 19}

\pagerange{\pageref{firstpage}--\pageref{lastpage}} \pubyear{2014}

\maketitle

\label{firstpage}

\begin{abstract}
  We present an analysis of the photometry and spectroscopy of the
  host galaxy of {\em Swift}-detected GRB\,080517. From our optical
  spectroscopy, we identify a redshift of $z=0.089\pm0.003$, based on
  strong emission lines, making this a rare example of a very local,
  low luminosity, long gamma ray burst. The galaxy is detected in the
  radio with a flux density of $S_{4.5\,GHz}=$0.22$\pm$0.04\,mJy - one
  of relatively few known GRB hosts with a securely measured radio
  flux. Both optical emission lines and a strong detection at 22$\mu$m
  suggest that the host galaxy is forming stars rapidly, with an
  inferred star formation rate $\sim16$\,M$_\odot$\,yr$^{-1}$ and a
  high dust obscuration (E$(B-V)>1$, based on sight-lines to the
  nebular emission regions). The presence of a companion galaxy within
  a projected distance of 25\,kpc, and almost identical in redshift,
  suggests that star formation may have been triggered by
  galaxy-galaxy interaction. However, fitting of the remarkably flat
  spectral energy distribution from the ultraviolet through to the
  infrared suggests that an older, 500\,Myr post-starburst stellar
  population is present along with the ongoing star formation. We
  conclude that the host galaxy of GRB\,080517 is a valuable addition
  to the still very small sample of well-studied local gamma-ray burst
  hosts.
\end{abstract}


\section{Introduction}

Long Gamma Ray Bursts (GRBs) are intense, relativistically beamed, bursts of
radiation, likely emitted during the collapse of a massive star at the
end of its life \citep{2006ApJ...637..914W}. As well as constraining the
end stages of the evolution for massive stars, they also mark out star
formation in the distant Universe, in galaxies often too small to observe
directly through their stellar emission or molecular gas 
\citep[e.g.][]{2012ApJ...754...46T}.  However, extrapolating from the detection of a single
stellar event (the burst) to their wider environment, and the
contribution of their hosts to the volume averaged cosmic star
formation rate \citep[e.g.][]{2012ApJ...744...95R}, is
challenging. Doing so relies on a good understanding of the stellar
populations and physical conditions that give rise to GRB events. 

This understanding has improved significantly over recent years.  A
number of studies now constrain the stellar properties of typical
GRB hosts \citep[e.g.][]{2009ApJ...691..182S,2010MNRAS.tmp..479S,2012ApJ...756..187H}, their
radio properties
\citep[e.g.][]{2012ApJ...755...85M,2010MNRAS.409L..74S,radiopaper,2014arXiv1407.4456P}
and behaviour in the far-infrared
\citep{2014arXiv1402.4006H,2014arXiv1406.2599S}. However these studies have also
demonstrated diversity within the population. GRB host galaxies range
from low mass, metal poor galaxies forming stars at a moderate rate
\citep[e.g.][]{2010AJ....140.1557L}, to more massive moderately dusty
but not extreme (SMG-like) starbursts such as the `dark' burst
population \citep{2013ApJ...778..128P,2013ApJ...778..172P}.

The challenge of understanding these sources has been complicated by
the high redshifts at which they typically occur.  The long GRB
redshift distribution peaks beyond $z=1$ \citep{2012ApJ...752...62J},
tracing both the rise in the volume-averaged star formation rate and
the decrease in typical metallicity - which may favour the formation
of GRB progenitors \citep[see e.g.][and references
  therein]{2012ApJ...744...95R}; local examples which can be studied
in detail are rare. Of long duration ($>$2s) bursts in the official
{\em Swift Space Telescope} GRB catalogue
table\footnote{http://swift.gsfc.nasa.gov/archive/grb\_table/}, only
three are listed as having $z<0.1$. A few other (pre-{\em Swift})
bursts are also known at low redshifts \citep[e.g. GRB\,980425 at
  $z=0.009$][]{1998Natur.395..670G}, but were detected by instruments
with quite different systematics and tend to be unusual systems. One
of the most recent studies, which exploited ALMA data, identified the
host of GRB\,980425 as a dwarf system with low dust content and
suggested that this is typical of GRB hosts as a whole
\citep{2014A&A...562A..70M}. However each low redshift host
investigated in detail has informed our understanding of the
population as a whole and proven to differ from the others
\citep[e.g.][]{2011ApJ...741...58W,2011MNRAS.411.2792S}. Low redshift
bursts include several which are sub-luminous, such as GRBs\,090825
and 031203
\citep{1998Natur.395..670G,2004ApJ...609L...5M,2004Natur.430..648S},
and others such as GRBs\,060505 and 060614 that were long bursts without
associated supernovae
\citep{2006Natur.444.1047F,2006Natur.444.1050D}. Cross-correlation
with local galaxy surveys (at $z<0.037$) has suggested that some low
redshift GRBs in the existing burst catalogues have yet to be
identified as such \citep{2007MNRAS.382L..21C} and hence opportunities
to study their properties in detail have been missed. Given the very
small sample, and the variation within it, it is important that we
continue to follow up the hosts of low redshift bursts and do not
allow a few examples to skew our perception of the population.

We have acquired new evidence suggesting that a previously overlooked
burst, GRB\,080517, and its host galaxy might prove a valuable
addition to the study of local gamma ray bursts. The WISE all-sky
survey \citep{2010AJ....140.1868W}, publically released in 2012, maps
the sky at 3-22\,$\mu$m. While the observations are relatively shallow
and most GRB hosts remain undetected or confused, we have identified
the host of GRB\,080517 as anomalous. Not only is an infrared-bright
source clearly detected coincident with the burst location, but it has
a sharply rising spectrum and is extremely luminous in the 22\,$\mu$m
W4 band, suggesting that it is a rather dusty galaxy, and likely at
low redshift.

In this paper, we present new photometry and spectroscopy of the
host of GRB\,080517, identifying its redshift as $z=0.09$. Compiling
archival data, we consider the spectral energy distribution (SED) of the
host galaxy, and also its larger scale environment, evaluating the
source as a low redshift example of a dusty GRB host galaxy.  In
section \ref{sec:initial} we discuss the initial identification of this
GRB and its properties. In section~\ref{sec:data} we
present new data on the host galaxy of this source. We present our optical 
photometry and spectroscopy of the
GRB host and a neighbouring companion in section \ref{sec:spec} and report a detection of
the GRB host at radio frequencies in section \ref{sec:radio}. In section \ref{sec:reassess} we reassess the
initial burst properties and its early evolution in the light of our new redshift information. 
In section \ref{sec:sed} we compile new and archival photometry to secure an analysis of
the spectral energy distribution, and in section \ref{sec:sfr} report constraints
on the host galaxy's star formation rate.
In section \ref{sec:disc} we discuss the properties of the host galaxy in the context
of other galaxy populations before presenting our conclusions
in section \ref{sec:conc}.

Throughout, magnitudes are presented in the AB system \citep{1983ApJ...266..713O} and
fluxes in $\mu$Jy unless otherwise specified. Where necessary, we use
a standard cosmology with $h_0$=70\,km\,s$^{-1}$\,Mpc$^{-1}$,
$\Omega_M=0.3$ and $\Omega_\Lambda$=0.7.

\section{Initial Observations}\label{sec:initial}

GRB\,080517 triggered the {\em Swift} Burst Alert Telescope (BAT) at
21:22:51 UT on 17th May 2008 as a flare with a measured T$_{90}$
(i.e. period during which 90\% of the burst energy was detected) of
65$\pm$27\,s, classifying the event as a long GRB. The X-ray
Telescope (XRT) identified a fading, uncatalogued point source and the
presence of a known optical source was noted within the X-Ray error
circle. The final enhanced XRT position, with uncertainty $1\farcs5$,
was 06h 48m 58.03s +50° 44′ 07.7′′ (J2000), coincident with the
optical source \citep{2008GCN..7742....1P}. The Galactic longitude and
latitude ($l=$165.369, $b=$20.301) correspond to a sight-line with
moderate dust extinction (A$_V$=0.25) from our own galaxy
\citep{2011ApJ...737..103S}.

Early observations with the Liverpool Telescope, starting 11 minutes
after the BAT trigger, did not detect an optical transient outside of the
known source \citep{2008GCN..7743....1S} and no further optical follow-up was undertaken - in
part due to the difficult proximity (within 50$^\circ$) of the Sun at the
time the burst triggered {\em Swift}. Both the lack of an optical afterglow and
analysis of the BAT spectrum suggested that the source might lie at
high redshifts \citep{2008GCN..7748....1M,2011ApJ...731..103X}, but
constraints on the X-ray spectrum precluded a high redshift fit to
the data \citep{2008GCN..7742....1P}. Association with the known, bright
optical
source would suggest a lower redshift for the burst, but it was not clear
whether this was the host galaxy or a star in chance alignment.

While the afterglow was not detected in the optical, the $\gamma$-ray
and X-ray emission was also relatively weak, with an early time flux
at 0.3-10\,keV of
2.52$^{+1.20}_{-0.75}\times10^{-10}$\,erg\,cm$^{-2}$\,s$^{-1}$,
measured in a 10\,s exposure at a mean photon arrival time of
T$_0$+133\,s \citep[based on analysis from][]{2009MNRAS.397.1177E}.
In the absence of a redshift for the host, the time-averaged X-ray
analysis also suggested the presence of an excess neutral hydrogen
column density of $3.0^{+2.1}_{-1.8} \times 10^{21}$ cm$^{-2}$ above
the Galactic value of $1.09\times10^{21}$\,cm$^{-2}$ \citep[where
  these are 90\% confidence intervals in analysis from the UK Swift
  Science Data Centre, ][]{2009MNRAS.397.1177E}. This represents an
excess in the X-ray inferred hydrogen column at the
$\sim$3\,$\sigma$ level.  {\em Swift} observations ended
approximately 20 hours after the initial trigger.

Initial observations for this source were therefore ambiguous, with
different elements of the data either suggesting a high redshift
solution (non-detection of the optical transient, BAT spectrum) or
appearing to preclude it (optical source association, X-ray spectrum), and
the excess extinction seen in the afterglow implying the presence of
dust either in the host galaxy or along the line of sight.  However
the burst's location, within 50$^\circ$ of the Sun at the time the
burst went off, precluded further early time studies, and the burst
has largely been ignored since. {\em Swift} has not observed this
location at any other time.

Given the presence of a relatively bright, $r_{AB}=17.73$, catalogued source
within the {\em Swift} XRT error circle, an obvious question arises: what
is the probability that this is a chance alignment rather than a genuine
host galaxy identification? Two main factors contribute to this determination. 
The surface density of galaxies observable at a given magnitude will
depend both on the properties of the galaxy population with redshift,
and with galactic latitude (which will govern the fraction of the sky
affected by foregrounds and crowding). To evaluate this, we have
studied the galaxy population in regions of the Sloan Digital Sky
Survey Data Release 10 \citep[SDSS DR10,][]{2014ApJS..211...17A} at comparable Galactic
latitude (within $\sim$5$^\circ$) and offset by 30-50$^\circ$ in
Galactic longitude.  The population identified by the SDSS photometric
pipeline as galaxies were selected in ten regions, each with a
diameter of 1$^\circ$, and their surface density evaluated as a
function of $r'$-band magnitude.

\begin{figure}
\begin{center}
\includegraphics[width=0.99\columnwidth]{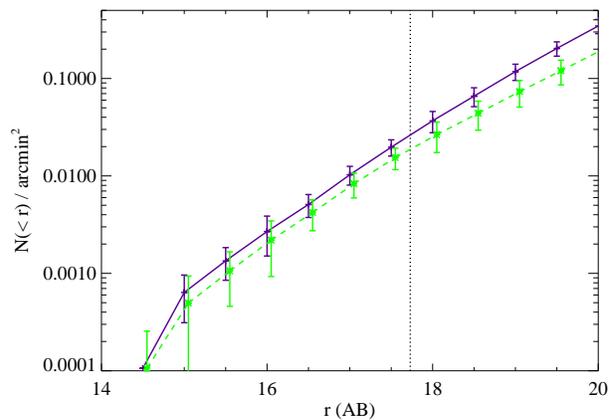}
\end{center}
\caption{The surface density of galaxies brighter than a given $r'$-band magnitude, at comparable galactic latitude to
GRB\,080517, based on photometric classification in the Sloan Digital Sky Survey. The solid line shows the surface density of all galaxies, with the standard deviation measured across ten 1$^\circ$-diameter fields. The dashed line shows the lower surface density of relatively blue galaxies likely to be star-forming. The dotted vertical line indicates the magnitude of the proposed host of GRB\,080517. \label{fig:coin}}
\end{figure}

As figure \ref{fig:coin} illustrates, the surface density of galaxies
comparable to the proposed host of GRB\,080517 is low, with
$0.028\pm0.006$ galaxies typically found per square
arcminute. Assuming that 
 the SDSS photometric classification is accurate, the
probability of finding a galaxy of this brightness within
3\,arcseconds (see figure \ref{fig:whtim}) of a given X-ray location is just 0.007\%. Taking into
account the 604 long bursts with X-ray localisations in the {\em Swift
  Space Telescope} GRB catalogue table, we would expect 0.04 chance
alignments amongst the entire long GRB population.

A further constraint arises from the nature of the GRB
itself. GRB\,080517 was a long burst, believed to be associated with a
core collapse progenitor, and so likely to be associated with recent
or ongoing star formation.  If we consider only those galaxies in SDSS
with the flat optical colours associated with ongoing star formation,
i.e. $|r'-i'|<0.5$, the surface density of galaxies drops still
further, to just $0.021\pm0.006$ galaxies per square arc minute, and a
predicted 0.03 chance alignments in the entire GRB sample.

As will be discussed below, the possible host galaxy of GRB\,080517 is strongly
star forming, and lies within 3 arcseconds of the burst location. Thus we propose its
identification as the burst host.

\section{Follow-Up Data}\label{sec:data}

\subsection{WHT Imaging}\label{sec:whtim}

We targetted the host of GRB\,080517 on the night of 2014 Feb 25
(i.e. 6 years post-burst) using the auxiliary-port camera, ACAM, on
the William Herschel Telescope (WHT). Photometric imaging was acquired
in the Sloan $g$, $r$ and $i$ bands, with an integration time of 180s
in each band.  Observations were carried out as part of programme
W/2014/9 (PI: Levan) and photometric data were
reduced and calibrated using standard {\sc IRAF} procedures.

\begin{figure}
\begin{center}
\includegraphics[width=0.99\columnwidth]{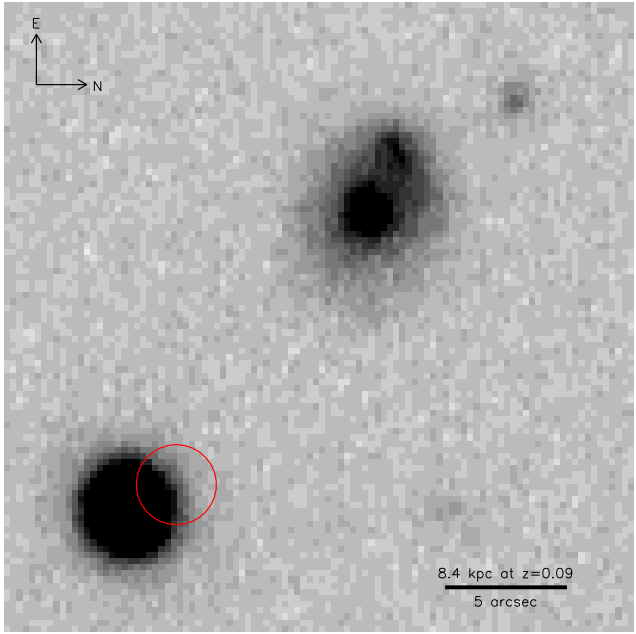}
\end{center}
\caption{The structure of the compact GRB host galaxy (lower left) and its near-neighbour (upper right) in the Sloan-$i$ band from our WHT imaging. The neighbour clearly has two cores within a more diffuse galaxy, and is likely to be undergoing a major merger. Both sources are at the same redshift (see section \ref{sec:spec}), the scale bar indicates physical distance at this redshift. The 1.5\,arcsecond 90\% confidence error circle from the {\em Swift} XRT detection of the burst is indicated in red. \label{fig:whtim}}
\end{figure}

As figure \ref{fig:whtim} demonstrates, the 1.2\,arcsecond seeing was
sufficient to determine the morphology of both the host galaxy and a
near neighbour, separated from it by 16 arcseconds.  While the GRB
host shows a relatively smooth, relaxed morphology, it is 
resolved in our imaging with a measured gaussian FWHM of
2.1\,arcseconds. Deconvolution with the seeing, as measured from
unresolved sources in the image, yields an estimated intrinsic size of
1.6\,arcseconds, or 2.7\,kpc at $z=0.09$ (see next section).

\begin{figure}
\begin{center}
\includegraphics[width=0.99\columnwidth]{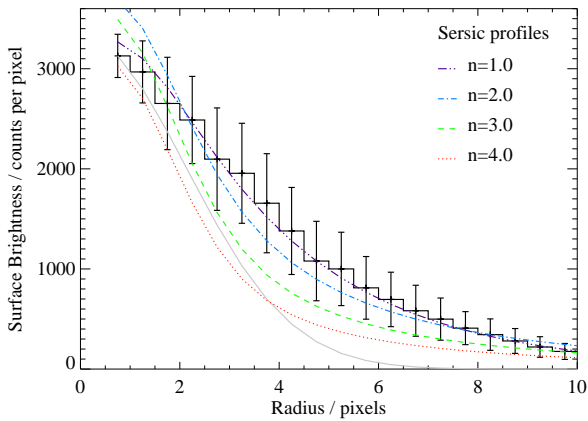}
\end{center}
\caption{The radial surface brightness profile of the GRB host galaxy in the Sloan-$i$ band from our WHT imaging. Sersic profiles have been convolved with the seeing  and overplotted for comparison. Given the large error bars - due to compact morphology relative to the pixel scale and seeing - a range of Sersic parameters provide a reasonable fit to the data. Normalising the profiles close to the centre suggests a Sersic index of $n\sim1.0-2.0$ may provide the best description of this galaxy's light profile. The gaussian seeing is shown as a solid line for comparison. \label{fig:sersic}}
\end{figure}

\begin{figure}
\begin{center}
\includegraphics[width=0.99\columnwidth]{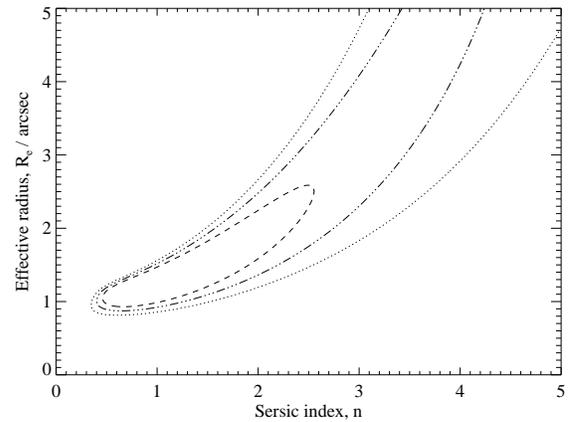}
\end{center}
\caption{The allowed regions of parameter space for a Sersic light profile, quantified by $\chi^2$-fitting against the data. Small effective radii ($<2''$) and low Sersic indices ($n\sim1-2$) are favoured by the data, but there are substantial degeneracies between these parameters. Contours are shown at 1, 2 and 3\,$\sigma$ confidence levels. \label{fig:sersic2}}
\end{figure}

While we are unable to distinguish clumpiness on sub-kiloparsec scales, the
host galaxy is sufficiently resolved in this new imaging to place
constraints on its radial light profile, although such constraints are
necessarily limited by the relatively large (0.253 arcsecond) pixels
relative to the seeing. In figures \ref{fig:sersic} and
\ref{fig:sersic2} we compare the radially averaged light profile of
the galaxy with Sersic profiles \citep[see][for definitions and
  discussion]{2005PASA...22..118G}, which have been convolved with the
seeing in the image. It is clear that a de Vaucouleurs law ($n=4$),
such as describes local giant elliptical galaxies would predict far
too steep a light profile. Allowing the effective radius and Sersic
parameter to vary simultaneously, the best fit to the data is found
for $n=1.5\pm1.0$ and $R_e=1.7\pm0.8''$.

\subsection{WHT Spectroscopy}\label{sec:spec}

We also obtained spectroscopic data from ACAM on the same night, using
the V400 grating and a total integration time of $4\times600$s,
producing a spectrum spanning 4000-9000\AA\ with a spectral resolution
measured from unblended arc lines of $\sim$18\AA\
($\sim$1000\,km\,s$^{-1}$). Both photometric and spectroscopic data
were reduced and calibrated using standard {\sc IRAF} procedures.
Absolute flux and wavelength calibration were achieved through
observations of a standard star field and arc lamps immediately
preceding the science data.

The slit was oriented at a position angle of 50$^\circ$, so as to
pass through the centres both of the GRB host and the bright neighbour, separated from it by
16$\arcsec$ measured along the 1$\farcs5$ slit.

\begin{figure*}
\begin{center}
\includegraphics[width=1.97\columnwidth]{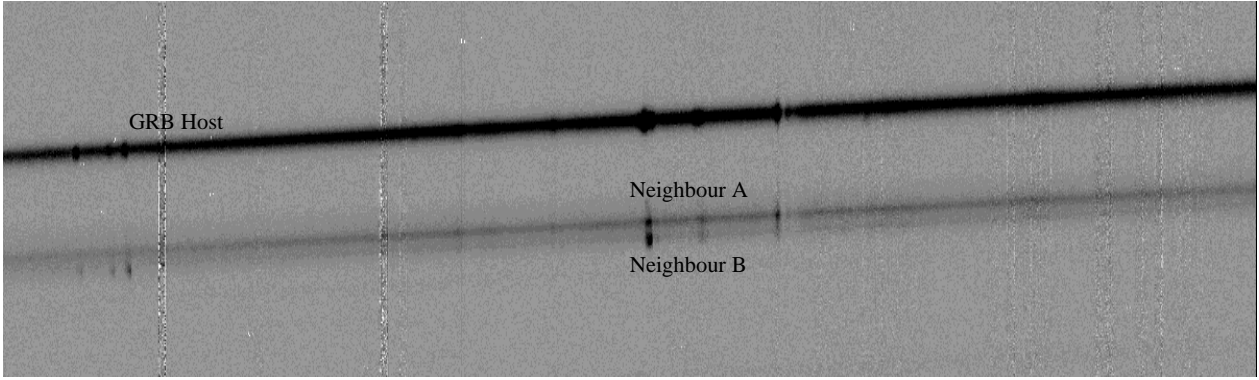}
\end{center}
\caption{Two dimensional spectrum observed for the GRB host galaxy (at top of the slit) and its neighbour, which divides into two components - A and B. All three components are strong optical line emitters, detected in multiple lines, at the same redshift to within the instrumental resolution. Spectra are aligned such that wavelength increases linearly towards the right of the frame. \label{fig:2dspec}}
\end{figure*}

\begin{figure*}
\begin{center}
\includegraphics[width=1.97\columnwidth]{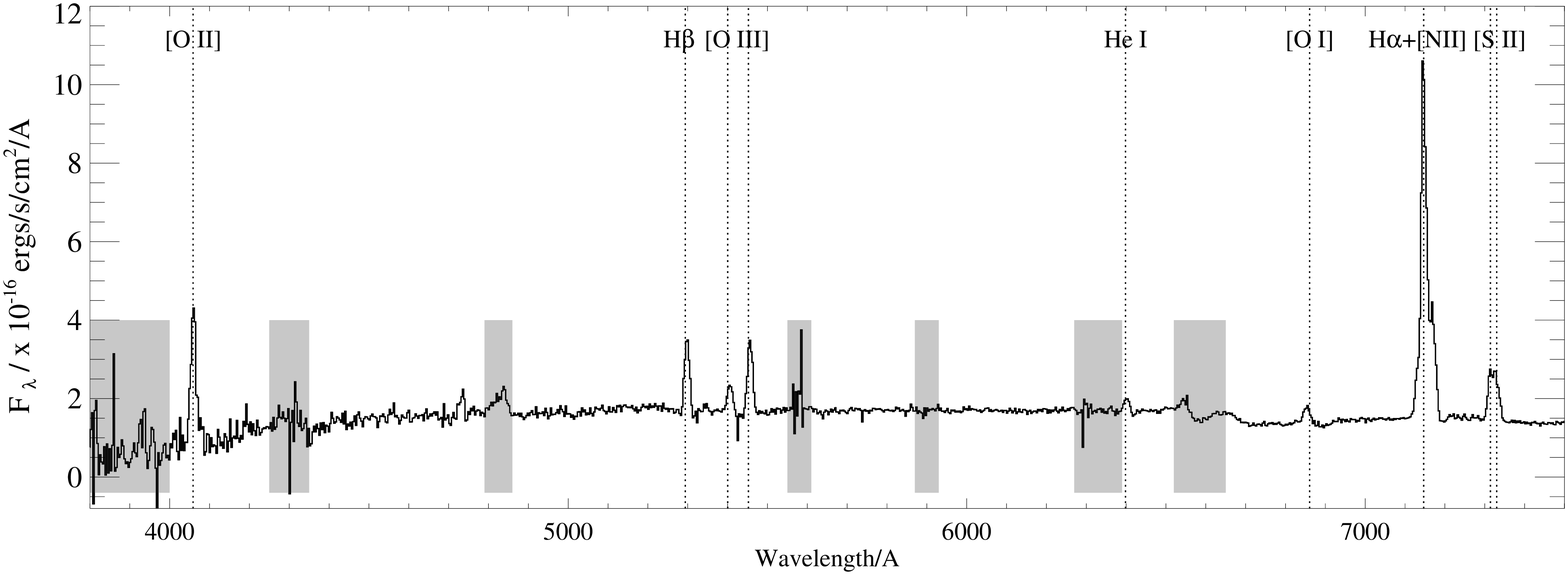}
\includegraphics[width=1.97\columnwidth]{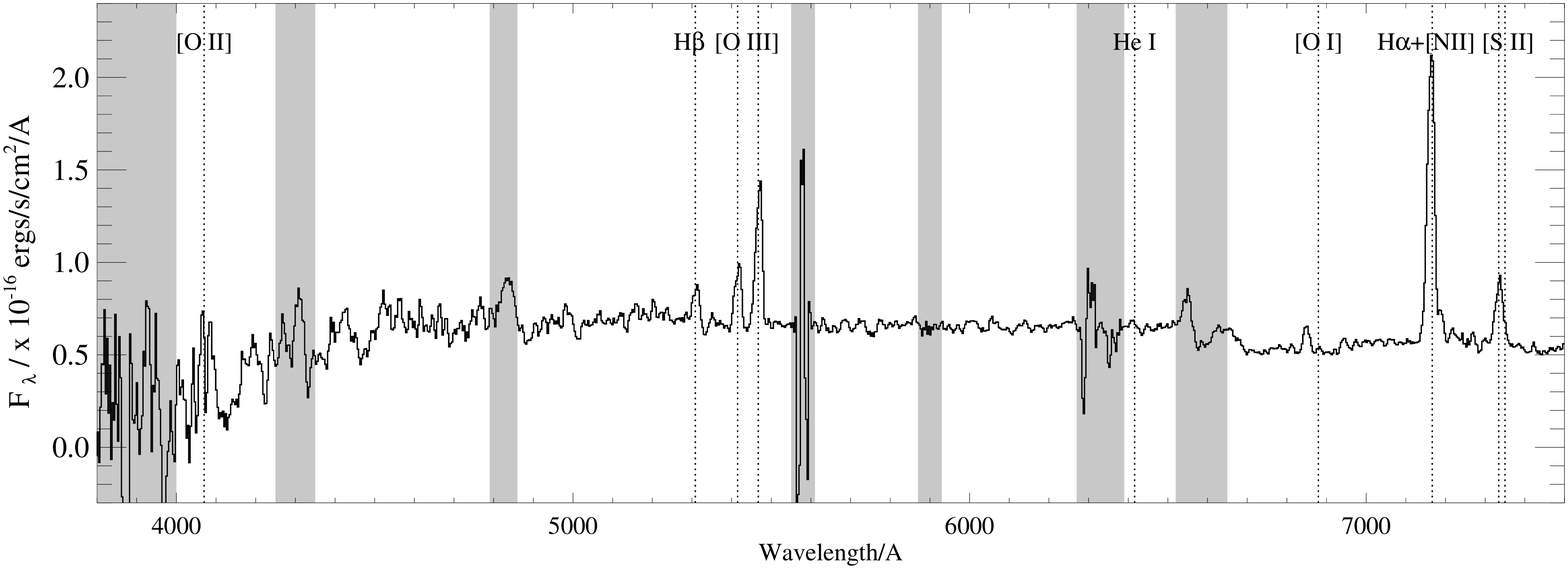}
\end{center}
\caption{The optical spectra of both the GRB host galaxy (upper plot) and its neighbour (lower).  Both sources show strong line emission, well detected continuum flux, and some evidence for a 4000\AA\ break. Wavelengths corresponding to [S\,{\sc II]}, [N\,{\sc II}], H$\alpha$, O\,{\sc I}, He\,{\sc I}, [O\,{\sc III}], H$\beta$ and [O\,{\sc II}] are indicated. Regions with relatively high noise caused by sky emission line subtraction residuals are indicated by shaded boxes. \label{fig:specall}}
\end{figure*}

Both the GRB host galaxy and its neighbour are detected at high signal
to noise in our spectroscopy. The latter clearly shows two components,
A and B. Of these, component A is the stronger continuum source, while
component B appears to show relatively stronger line emission (see
figure \ref{fig:2dspec}).  In table \ref{tab:lines} we provide the
relative emission line strengths of each source (also shown
graphically in figure \ref{fig:specall}). Line equivalent widths are
presented in the observed frame. We make no adjustment for slit losses
since it is difficult to reconstruct precisely where on the object the
1.5 arcsec wide slit was placed, and harder still to estimate whether
line ratios in the regions of the galaxy outside the slit are
comparable to those in observed regions. The measured redshift for the
host galaxy is $z=0.089\pm0.003$ and for the neighbour
$z=0.091\pm0.003$ (for both components). The uncertainty, estimated by
cross correlation against a template starburst spectrum, comprises
instrumental resolution effects, the effects of blending on many of
the lines and uncertainty due to small shifts in velocity offset
between different emission lines.

While we adopt the cross-correlation redshifts and conservative
associated uncertainties for our analysis, we also consider the
redshift derived from the observed wavelength of individual emission
lines. Fitting gaussian profiles to the unblended H$\beta$ and
[O\,III] and the strong, but somewhat blended H$\alpha$ lines, we
derive redshifts $z=0.0903\pm0.0006$, $0.0925\pm0.0006$ and
$0.0930\pm0.0003$ for the GRB host and components A and B of the
companion respectively, where the error now represents the scatter
between individual line centroids on each source rather than including
other uncertainties.  These imply velocity offsets of $\Delta v =
150\pm155$\,km\,s$^{-1}$ (i.e. no significant offset) between the two
components of the companion and $\Delta v = 576\pm155$\,km\,s$^{-1}$
between the host and the companion.

This velocity offset places the galaxy pair just outside (although
within one standard deviation of) the criteria used to select galaxy
pairs in the SDSS by \citet{2008AJ....135.1877E}, who placed a cutoff
for their sample at $\Delta v = 500$\,km\,s$^{-1}$. Those authors
recognise however that this cutoff requires a trade-off between
contamination and completeness, with genuine pairs observed out to
$\Delta v \sim 600$\,km\,s$^{-1}$ separations
\citep{2008AJ....135.1877E,2000ApJ...536..153P}. \citet{2008AJ....135.1877E}
identified an enhancement in star formation rate for pairs with
projected separations $<30-40$\,kpc, a criterion easily satisifed by
the companion in this case (16$''$ represents $\sim27$\,kpc at this
redshift), suggesting that the star formation observed in both host
and companion is likely influenced by their proximity.

\begin{figure}
\begin{center}
\includegraphics[width=0.98\columnwidth]{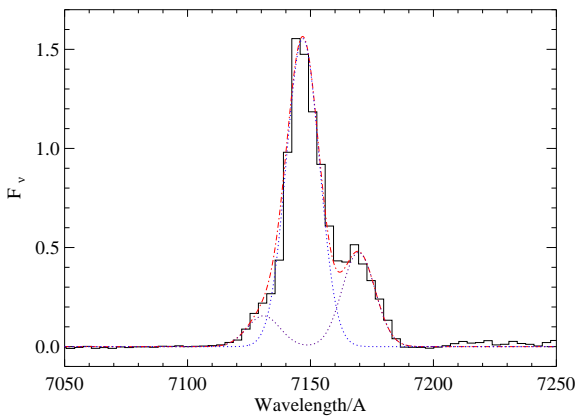}
\end{center}
\caption{The spectral region containing H$\alpha$ and the [N\,{\sc II}] doublet. All three lines are consistent with the being unresolved at the instrumental FWHM. The relative strength of the doublet lines is consistent with that predicted from the electron transition probabilities. \label{fig:specha}}
\end{figure}

\begin{figure}
\begin{center}
\includegraphics[width=0.98\columnwidth]{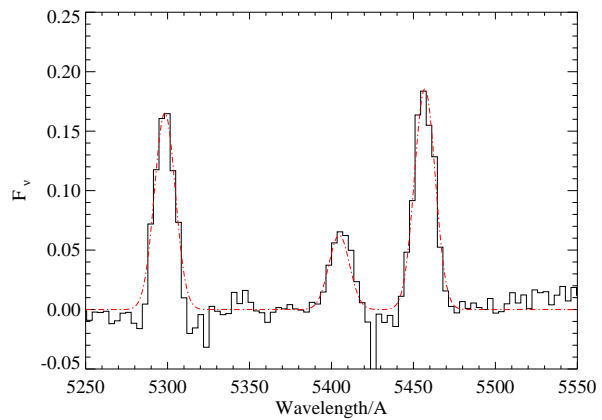}
\end{center}
\caption{The spectral region containing H$\beta$ and the [O\,{\sc III}] doublet.\label{fig:spechb}}
\end{figure}

In figures \ref{fig:specha} and \ref{fig:spechb}, we present the
spectral regions in the GRB host galaxy associated with line ratios
used to classify an ionising spectrum (see section \ref{sec:disc}). At
this spectral resolution, H$_\alpha$ is blended with the [N\,{\sc II}]
doublet, and a fit to the three lines must be obtained simultaneously
in order to measure their line strengths. With the exception of close
doublets (i.e. [O\,{\sc II}], [S\,{\sc II}]) the other
lines in the spectrum are all comparitively unblended, and all lines
are consistent with being essentially unresolved at the instrumental
resolution.

\begin{table}
\begin{tabular}{lccc}
Line  &  Host  & Neighbour A & B \\
\hline\hline
 \lbrack O\,{\sc II}\rbrack  3726      &  61.9 $\pm$ 8.0  &                 &                \\
 \lbrack O\,{\sc II}\rbrack  3729      &  31.0 $\pm$ 4.0  &                 &                \\
H$\gamma$                 &   6.5 $\pm$ 0.4  &                 &                \\
H$\beta$                  &  16.8 $\pm$ 1.0  &  3.2  $\pm$ 1.3 &   22 $\pm$ 9   \\
 \lbrack O\,{\sc III}\rbrack  4959     &   5.8 $\pm$ 0.3  &  4.1  $\pm$ 1.7 &   27 $\pm$ 11   \\
 \lbrack O\,{\sc III}\rbrack  5007     &  17.1 $\pm$ 1.0  & 12.4  $\pm$ 5.0 &   77 $\pm$ 32   \\
He I 5875                 &   2.8 $\pm$ 0.1  &                 &                 \\
 \lbrack O\,{\sc I}\rbrack  6300       &   6.1 $\pm$ 0.2  &                 &                 \\
 \lbrack N\,{\sc II}\rbrack 6548       &  10.3 $\pm$ 0.2  &  1.6  $\pm$ 0.1 &   4.1 $\pm$ 0.4 \\
H$\alpha$                 & 103.1 $\pm$ 2.4  & 31.8  $\pm$ 1.5 &  116  $\pm$ 12  \\
 \lbrack N\,{\sc II}\rbrack  6583      &  31.5 $\pm$ 0.7  &  4.9  $\pm$ 0.2 &  12.5 $\pm$ 1.2 \\
 \lbrack S\,{\sc II}\rbrack  6716      &  13.3 $\pm$ 0.5  &                 &                 \\
 \lbrack S\,{\sc II}\rbrack  6730      &  13.0 $\pm$ 0.5  &                 &                 \\
\end{tabular}
\caption{Line strengths measured for the target objects. All measures are given as observed-frame equivalent widths in Angstroms. Measurement of weak lines is not attempted in the fainter neighbour, and it is impossible to isolate the two components in the [O\,{\sc II}] line.\label{tab:lines}}
\end{table}

\begin{figure}
\begin{center}
\includegraphics[width=0.98\columnwidth]{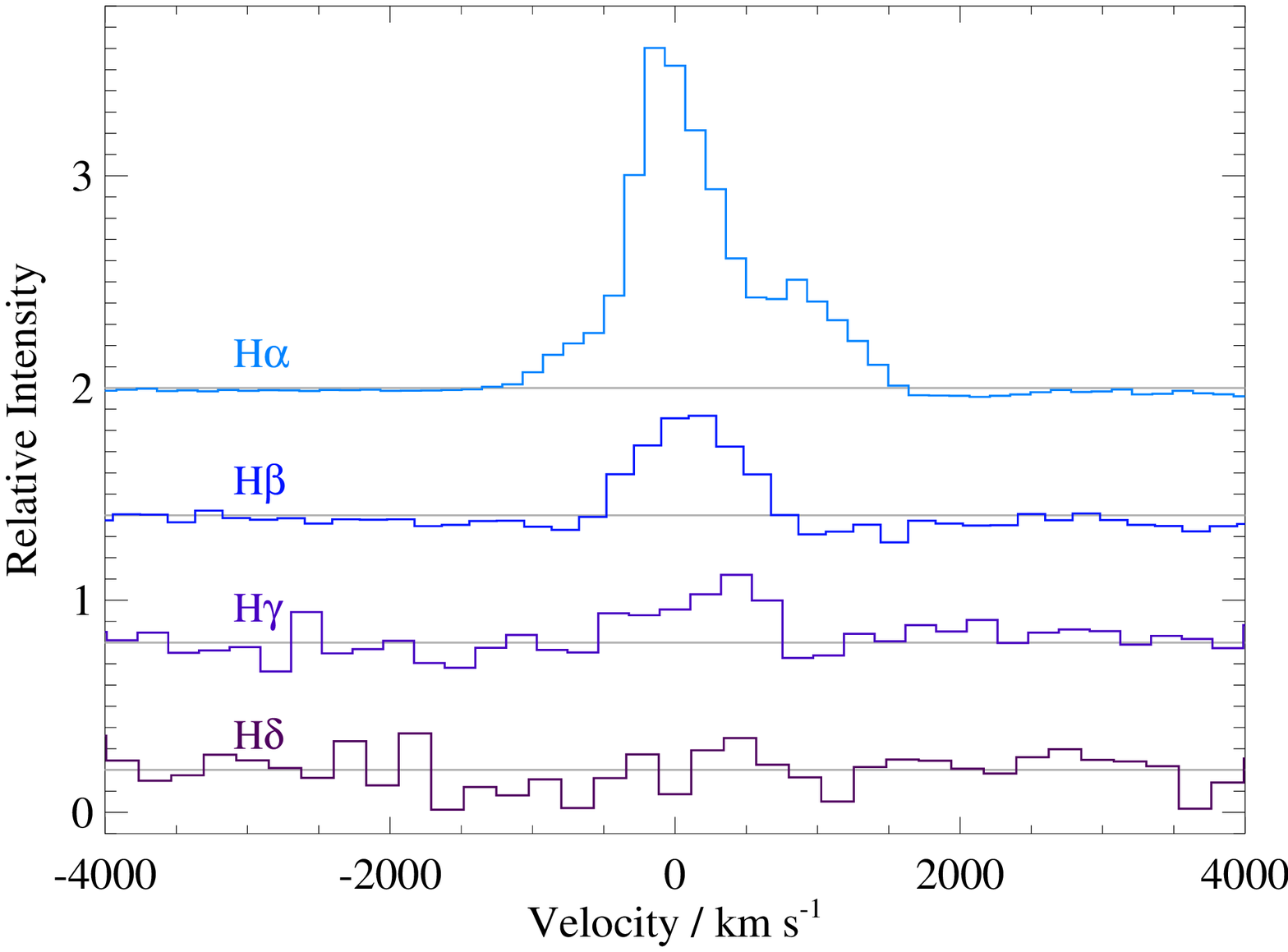}
\end{center}
\caption{A comparison of the Balmer series emission lines in the GRB host galaxy, scaled by the appropriate line ratios for Case B H\,{\sc I} recombination (T=$10^4$\,K, low density, Osterbrock \& Ferland 2006), such that in the absence of dust, all peaks would be expected to match H$\beta$ in intensity. The line intensities are offset from zero for clarity. Even accounting for blending with the [N\,{\sc II}] doublet, H$\alpha$ still shows a relative excess, consistent with the blue-wards lines being attenuated by a dusty line of sight. \label{fig:specratio}}
\end{figure}

The relative intensity of the nebular emission in hydrogen Balmer lines allows us to make an estimate of the dust extinction in the actively star forming region of the GRB host galaxy. The flux in H$\alpha$ is expected to scale relative to H$\beta$ by a ratio determined by the temperature and electron density of the emitting region. Standard assumptions for these parameters in HII regions (T=10,000\,K, low density limit) yields the widely applied expected ratio of 2.87 \citep[see ][]{2006agna.book.....O}. In figure \ref{fig:specratio}, we scale the line fluxes from the GRB host galaxy accordingly, such that, for Case B reionisation, we would expect each line to have the same relative intensity as H$\beta$. It is clear that the Balmer series shows a decrement in the later lines, most likely attributable to dust.  Given a Calzetti-like dust law \citep[$R_V=4.05$,][]{2000ApJ...533..682C}, the measured H$\alpha$/H$\beta$ ratio implies an extinction of flux from the nebular emission region of the GRB host galaxy of E($B-V$)=1.2. 

Of course, uncertainty arises in whether the HII region parameters and extinction law adopted are indeed appropriate for GRB hosts galaxies. \citet{2011MNRAS.414.2793W} explored a grid of temperatures and dust extinction laws for fitting the H\,{\sc I} Balmer series in example spectra and suggested that in the case of the GRB\,060218 host a higher temperature and steeper extinction law (T=15,000\,K, $R_V=4.5$) might be appropriate, while the host of GRB\,100316D is best fit with T$\sim7500$\,K and $R_V=3.5$. The Calzetti dust law lies between that inferred from these two examples, as does our adopted temperature. More detailed spectroscopy (with fainter Balmer lines, and ideally also the He II recombination series) would be required to reach a tighter constraint.

We also obtain a tentative spectroscopic redshift for an unrelated
galaxy falling on the long slit.  The galaxy, located at RA and
Declination 06$^h$48$^m$59.756$^2$ +50$^\circ$44$'$23.50$''$ (J2000),
lies at $z=0.56$, based on identification of an emission feature as
the [O\,{\sc II}] 3727\AA\ doublet.

\subsection{Radio Observations}\label{sec:radio}

The low redshift confirmed for this GRB host makes it an ideal candidate
for radio observation.  The majority of radio observations of GRB
hosts to date have resulted in non-detections, implying star formation
rates that do not significantly exceed their UV-optical estimates
\citep[e.g.][]{2012ApJ...755...85M,2010MNRAS.409L..74S,radiopaper}. However,
some fraction of GRB hosts appear to be luminous in the
submillimetre-radio \citep{2003ApJ...588...99B,2004MNRAS.352.1073T},
particularly amongst those that show evidence for strong dust
extinction \citep[i.e. dark bursts,][]{2013ApJ...778..172P,2014arXiv1407.4456P}.

Radio observations of the GRB\,080517 host galaxy were performed with
the Westerbork Synthesis Radio Telescope (WSRT) at 4.8\,GHz on 2014
May 2 and May 3~UT, i.e. almost 6~years after
the gamma-ray trigger. We used the Multi Frequency Front Ends
\citep{tan1991} in combination with the IVC+DZB back end in continuum
mode, with a bandwidth of 8x20 MHz. Gain and phase calibrations were
performed with the calibrator 3C\,147. The data were analyzed using
the Multichannel Image Reconstruction Image Analysis and Display
\citep[{\sc MIRIAD};][]{1995ASPC...77..433S} package.

\begin{figure}
\begin{center}
\includegraphics[width=0.95\columnwidth]{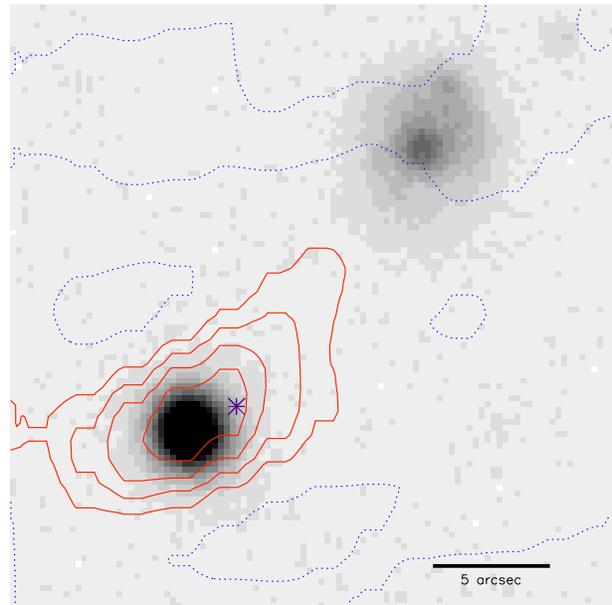}
\end{center}
\caption{The 4.8\,GHz radio flux measured at the WSRT (contours), overlaying the compact GRB host galaxy (lower left) and its near-neighbour (upper right) in the Sloan-$r$ band (greyscale). The contours indicate levels of zero flux (dotted) and +2, 3, 4 and 5\,$\sigma$. There are no signals below $-2$\,$\sigma$ in this region. The burst location is indicated with a cross. \label{fig:radim}}
\end{figure}

Both observations resulted in a detection of a source at the position
of GRB\,080517, with consistent flux densities. We have measured the
flux density in an image of the combined data set as
$S_{4.8\,GHz}=$0.22$\pm$0.04\,mJy. The detection is consistent with a
point source, in observations with a synthesised beam of
$14.2\times5.3''$ as shown in figure \ref{fig:radim}. No significant
detection is made of the neighbour galaxy - somewhat surprisingly
given its high inferred star formation rate (based on H$\alpha$
emission, see section \ref{sec:disc}).

\section{Reassessing GRB\,080517}\label{sec:reassess}

Given the identification of a redshift for the host galaxy of
GRB\,080517, we are able to reassess the properties of the burst and
its immediate afterglow in the context of an accurate distance (and
thus luminosity) measurement, allowing for more meaningful comparison with
the rest of the GRB population.

\subsection{Burst properties}\label{sec:grb}

At $z=0.09$ the inferred isotropic equivalent energy of GRB\,080517 is
only $E_{iso} = (1.03 \pm 0.21) \times 10^{49}$ ergs, while its 10
hour X-ray luminosity is $L_X \sim 10^{42}$ ergs s$^{-1}$. Both of
these values lie orders of magnitude below the expectations for most
GRBs, which have characteristic values of $E_{iso} \sim 10^{52-54}$
ergs \citep{2008ApJ...680..531K,2009ApJ...693.1484C} and $L_X \sim
10^{45-47}$ ergs s$^{-1}$ \citep{2006ApJ...642..389N}. These
properties mark GRB\,080517 as a member of the observationally rare
population of low luminosity GRBs (LLGRBs). Only a handful of such low
luminosity events have been identified in the past decade, all of
which have been relatively local (given the difficulty in observing
low luminosity bursts beyond $z\sim0.1$).  These include the well
studied GRB-supernova pairs GRB\,980425/SN~1998bw \citep{1998Natur.395..670G},
GRB\,031203/SN~2003lw \citep{2004ApJ...609L...5M,2004Natur.430..648S}, GRB
060218/SN~2006aj \citep{2006Natur.442.1011P} and GRB\,100316D/SN~2010bh
\citep{2011MNRAS.411.2792S,2011ApJ...740...41C}, and the enigmatic GRBs 060505
and 060614, where associated SNe have been ruled out to deep limits,
and whose origin remains mysterious \citep{2006Natur.444.1047F,2006Natur.444.1050D}. They
may be low luminosity events akin to those above, but where the SNe is
absent \citep[e.g.][]{2006Natur.444.1053G}, or they could be GRBs with a similar
physical origin to the short-GRBs \citep[most likely NS-NS mergers based on
recent observations, ][]{2006Natur.444.1044G,2013Natur.500..547T}, in which case their
luminosities would be more typical of their population. GRB 080517
adds a further example to these local, low luminosity events. Its
highly star forming host galaxy (as discussed later) is perhaps most in keeping with the
expectations of long-GRBs, although its high stellar mass and
metallicity would be unusual at such low redshift
\citep{2013ApJ...774..119G}.

The prompt emission from GRB\,080517, as reported by the {\em
  Swift}/BAT instrument, shows a single `fast rise, exponential decay'
\citep[FRED][]{1995PASP..107.1145F} lightcurve, albeit at low signal
to noise. In this respect, its profile is similar to that of low
luminosity GRBs\,031203 and 980425 \citep{2007ApJ...654..385K},
although the profile is not unusual amongst GRBs more generally
\citep{1996ApJ...473..998F}.

Interestingly, within the low luminosity population there appears to be a good deal
of internal diversity. The {\em Swift}-identified low luminosity
events to date -- GRB\,060218 \citep{2006Natur.442.1008C} and
GRB\,100316D \citep{2011MNRAS.411.2792S} -- appear to be of extremely
long duration (2000\,s in the case of GRB\,060218) with extremely
smooth light curves. They are also very soft events in which the X-ray
emission exceeds that in the $\gamma-$ray (so called X-ray
Flashes). In contrast, the pre-{\em Swift} events (GRB\,980425 and
GRB\,031203) appear to be much closer in prompt properties to
classical GRBs, exhibiting shorter durations (tens of seconds) and
relatively hard $\gamma$-ray spectra. Although there has been 
some suggestions that GRB 031203 was a softer X-ray event, integrated over
longer time periods, this is based on inferences from its X-ray echo \citep{2004ApJ...603L...5V}, and
favour a soft component arising after the initial burst \citep{2004ApJ...605L.101W}. In retrospect this
is most likely an X-ray flare, as are commonly seen in {\em Swift} X-ray
afterglows \citep{2006ApJ...636..967W,2006ApJ...642..389N}. Hence we consider the $E_p$ measured
via {\em INTEGRAL} as likely indicative of the true burst $E_p$ \citep{2004Natur.430..646S}. 
In this case, both GRB 980425 and GRB 031203 lie well
away from the correlation between the peak of the $\nu F_{\nu}$
spectrum ($E_p$) and $E_{iso}$
\citep{2002A&A...390...81A,2006MNRAS.372..233A}. 

\begin{figure}
\begin{center}
\includegraphics[width=1.05\columnwidth]{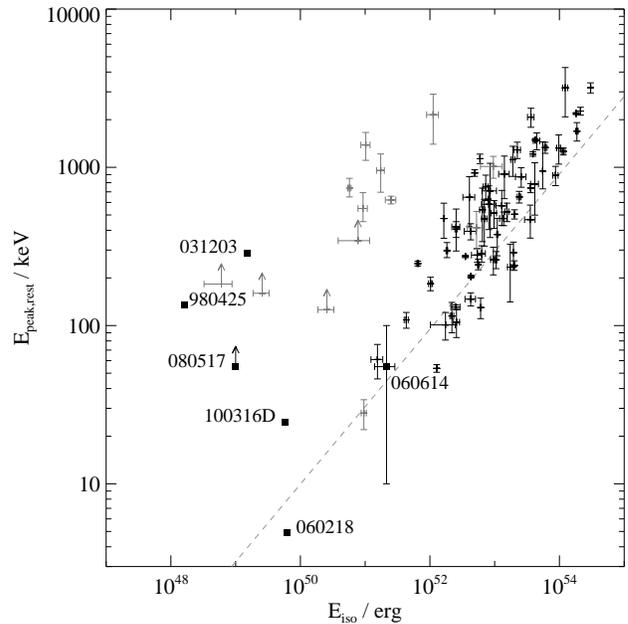}
\end{center}
\caption{The location of GRB\,080517 on the $E_p$ -- $E_{iso}$ (Amati) relation, given its
 redshift of $z=0.09$. Black points indicate long GRBs, while those in grey are the short GRB population. GRB\,080517 and 
previously identified low luminosity bursts are labelled. The burst lies in an unusual region of parameter space, well below
 the commonly seen relation for GRBs, placing it in the class of low redshift, low luminosity bursts.  \label{fig:amati_rel}}
\end{figure}

GRB\,080517 appears to have much more in common with these pre-Swift
events, with a $T_{90} = 65\pm27$s and a hard photon index of $\Gamma
\sim 1.5$. While $E_p$ is difficult to directly constrain with the
limited BAT bandpass, the Bayesian method of
\citet{2007ApJ...663..407B} suggests that $E_p > 55$ keV, making
GRB\,080517 a significant outlier in the $E_p$ -- $E_{iso}$ relation,
with a similar location to GRB\,980425 and GRB\,031203 (see
figure~\ref{fig:amati_rel}). Its recovery, some 6 years after the
initial detection implies that other, similar, low luminosity events
may be present within the {\em Swift} catalog, since a significant
number of bursts have not been followed in depth due to observational
constraints. However, GRB\,080517 was unusual in having a bright
catalogued source within its error box - a relatively rare occurrence.
In this context, it should be noted that the host of GRB\,080517 is
relatively luminous for a GRB host. By contrast, the hosts of GRB\,980425
and GRB\,060218 would have had observed magnitudes of $r\approx$20 and
$r\approx$22.5 at $z=0.1$ and so would not be readily cataloged in DSS
and similar survey observations, suggesting that the presence of a catalogued
source is not necessarily a good indicator of event frequency.

\subsection{Afterglow reanalysis}\label{sec:afterglow}

Making use of data analysis tools available from the UK Swift Data
Centre\footnote{http://www.swift.ac.uk \citep{2009MNRAS.397.1177E}},
specifying the burst redshift as matching that determined for the
host, we have reanalysed the Swift XRT afterglow spectrum and early
time series data.

Allowing for absorption at the host redshift only slightly modifies
the hydrogen column density required to fit the burst X-ray spectrum
observed by {\em Swift}. The effect on the late time spectrum (mean
photon arrival time = T0+25863, where T0 is the {\em Swift} trigger
time) is negligible, not modifying the required intrinsic absorption
from the N$_H=6^{+4}_{-5}\times10^{21}$\,cm$^{-2}$ in excess of the
estimated Galactic absorption estimated with the absorber at $z=0$
(where the errors on N$_H$ from Swift are 90\% confidence rather than
1$\sigma$ intervals).  The early time PC-mode data, with a mean photon
arrival of T0+9559s, yields a lower (but consistent) estimated
intrinsic absorption
(N$_H=3.4^{+2.4}_{-2.0}\times10^{21}$\,cm$^{-2}$), and a photon index
of 1.9$\pm$0.4.

\begin{figure}
\begin{center}
\includegraphics[width=0.99\columnwidth]{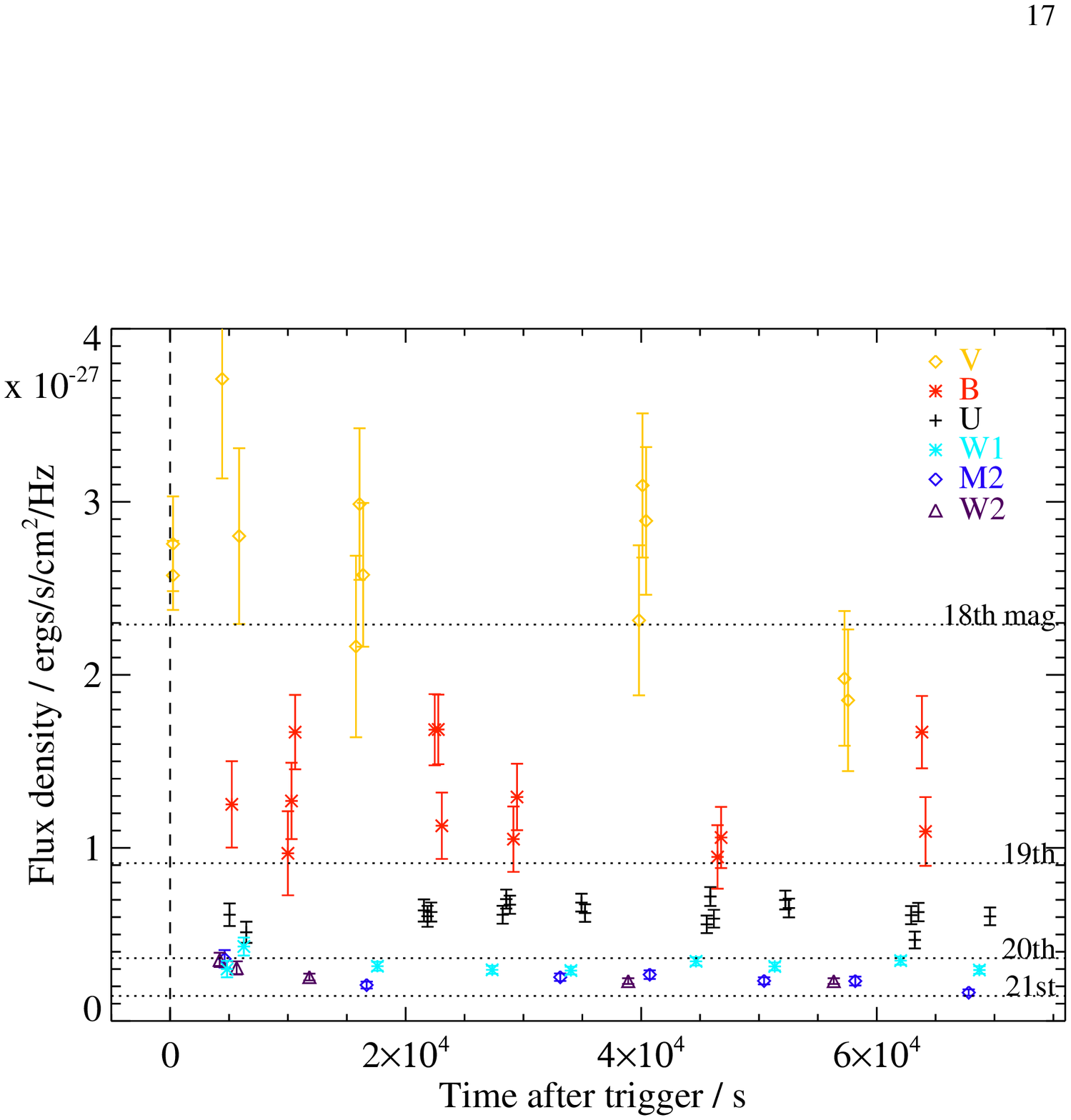}
\end{center}
\caption{Early time optical and ultraviolet photometry from {\em Swift}/UVOT. The flux density in the $B$ and $U$-band and the three ultraviolet wavebands shows little evidence of variation. There is a hint of declining flux in the $V$-band, but given the large photometric errors in this band, any decline is difficult to constrain with any accuracy. \label{fig:uvotearly}}
\end{figure}

Optical/ultraviolet imaging was also obtained by the {\em Swift}/UVOT
instrument, from first acquiring the field to the end of observations
at T0+19hrs. Data were observed in 6 bands ($V$, $B$, $U$, $UVW1$,
$UVM2$ and $UVW2$), and photometric imaging was obtained in each band at intervals
throughout this early period. As figure \ref{fig:uvotearly}
demonstrates, these is little evidence of temporal variation in five
of the six bands. Each is consistent with a constant flux. There is
a hint of declining flux in the last observations taken in the reddest, $V$, band
but the substantial uncertainties on this data preclude a firm
identification of afterglow flux. No other band shows a comparable
decline during the observation interval.

\begin{figure}
\begin{center}
\includegraphics[width=0.99\columnwidth]{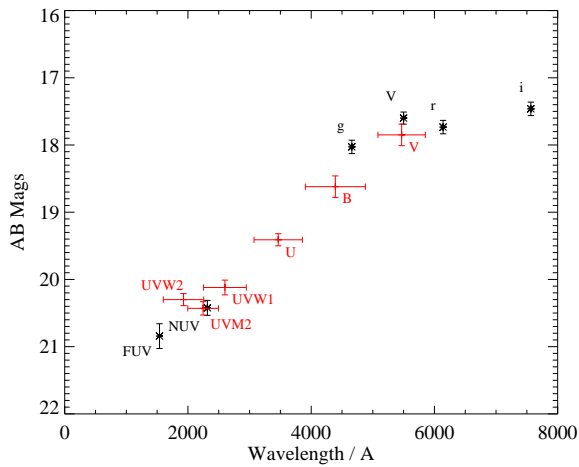}
\end{center}
\caption{Comparison of averaged early time optical and ultraviolet photometry from {\em Swift}/UVOT (red, crosses), with the late time observations of the host from other sources (see section \ref{sec:sed}). The UVOT observations are averaged from T0+0hr to T0+19hr. The late time observations are taken at several years post-burst. Nonetheless, the UVOT observations are consistent (within the photometric errors) with the host galaxy data, suggesting that any afterglow was below the UVOT detection limit. \label{fig:uvotcomp}}
\end{figure}

In figure \ref{fig:uvotcomp}, we compare this early time UVOT data,
now integrated across the 19 hour observation assuming no temporal
variation, with the late time host galaxy data described in sections
\ref{sec:whtim} and \ref{sec:sed}. While observations in $U$ and $B$
are not available at late times, the measured flux in the early time
integrated $V$ band and near-ultraviolet bands are consistent with
that in late time observations of the host galaxy.  With the exception
of possible variation the $V$-band data, no afterglow is detected
within the photometric errors, suggesting that any optical supernova
was at or below the UVOT detection limit.  Taking the 1$\sigma$ upper
limit on the early time photometry, and subtracting off the late time
galaxy flux (see below), we constrain the optical afterglow to
$F_\lambda<2\times10^{-17}$\,ergs\,s$^{-1}$\,\AA$^{-1}$\ at T0+4000s, measured
at 5500\AA. 
The {\em Swift} detected X-ray flux at the same epoch
(T0$\sim5227^{+1471}_{-1016}$s), was
$(3.1\pm0.8)\times10^{-13}$\,ergs\,s$^{-1}$\,cm$^{-2}$ in the 0.3-10\,keV
band. Comparing these yields a limit on X-ray to optical ratio,
$\beta_{OX}<1.0$. 

Finally, we have reexamined the early time data from the Liverpool
Telescope (LT) observations. Smith et al (2008, GCN\,7743) reported
limits for non-detection of an optical transient, based on the
assumption that it was not coincident with the known source. LT
observations commenced at 21:34:05 UT, 674s after the burst trigger.
The data comprised imaging of the field in SDSS-$r'$, $i'$ and $z'$
bands, with 120s individual exposures, the last of which ended at
T0+2290s.  We confirm that there is no evidence for an early-time
excess due to afterglow flux in this source, either in subtractions
against our late-time WHT imaging or in pairwise subtraction of
early-time exposures. These data provide a relatively weak constraint,
but at an early time, enabling us to limit the flux at T0+900s to
$F_\lambda<2\times10^{-16}$\,ergs\,s$^{-1}$\,\AA\ at 7500\AA.
Comparing to the {\em Swift}-detected X-ray flux at the same epoch, we
determine an identical limit to that from the {\em Swift} optical data
alone -- $\beta_{OX}<1.0$.

These limits are formally too weak to satisfy the $\beta_{\rm{OX}}$
criterion that is applied to select dark bursts
\citep{2004ApJ...617L..21J,2009ApJ...699.1087V}. Thus the
non-detection of the afterglow is consistent with either `dark' or
`normal' interpretation. However, we note that this burst shows the
high column and red, dusty host more common amongst the dark
population \citep[e.g.][]{2014arXiv1402.4006H,2013ApJ...778..128P}.

\begin{figure}
\begin{center}
\includegraphics[width=0.99\columnwidth]{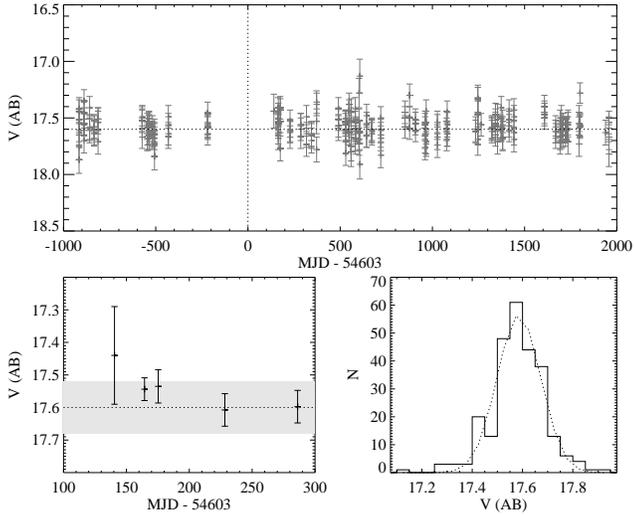}
\end{center}
\caption{Optical time-series data from the Catalina Real-time
  Transient Survey for the host galaxy of GRB 0801517. There is no
  evidence for strong variability in the host galaxy, allowing the time
  series data to be combined to determine a precise magnitude for the
  object. In the lower left panel we consider the data closest to the
  burst data (MJD=54603) in detail, averaging the data points on each
  night on which the host was observed. While the first three data
  points lie above the mean measured magnitude, they are within one
  standard deviation of the time-series mean (shown with shaded
  region).\label{fig:catalina}}
\end{figure}

Additional $V$-band time series data for this target exists in the
second data release of the Catalina Real-time Transient Survey
\citep[CRTS,][]{2009ApJ...696..870D}. Both the GRB host and neighbour
are well detected. Unfortunately the burst itself occured during a
hiatus in CRTS observing, with no data available until 140 days after
the GRB trigger (likely due to Sun avoidance). However, as figure
\ref{fig:catalina} demonstrates, the time series data of the GRB host
shows no strong evidence for variability (although a few photometric
points are outliers), and allows a precise measurement of the optical
magnitude of the host galaxy, $V_{AB}=17.60\pm0.08$. We also
investigate the late time optical afterglow, and find no statistically
significant evidence for an excess over the host galaxy flux at
T+160 days.

\section{The Spectral Energy Distribution}\label{sec:sed}

Our WHT optical imaging in the SDSS $g$, $r$, and $i$ bands (described
in section \ref{sec:whtim}) is supplemented by extensive archival data
on these relatively bright sources.

In addition to the $V$-band data from the Catalina Real-time Transient
Survey described in section \ref{sec:afterglow}, we compile archival
data in the ultraviolet from the {\em GALEX}
\citep[GR6,][]{2003SPIE.4854..336M} survey and in the near-infrared
from the Two Micron All-Sky Survey
\citep[2MASS,][]{2006AJ....131.1163S} as well as the {\em Wide-Field
  Infrared Survey Explorer} \citep[WISE,][]{2010AJ....140.1868W}. Both
the GRB host and its merging neighbour are detected in the majority of
bands from the near-ultraviolet (NUV, 0.15$\mu$m) through to the
mid-infrared (W4, 22$\mu$m). While flux from the neighbour is clearly
dominated by two principle components in our WHT imaging, these are
blended in the remaining data, and we do not attempt to gauge the
relative contribution of the two components. Photometry was measured
on the images at the source locations, and checked against catalog
magnitudes where these were available.

For the W3 and W4 bands (at 12 and 22$\mu$m) where the host and its
neighbour are blended in the imaging data, we use magnitudes derived
from the `ALLWISE' catalogue values (corrected to AB magnitudes)
rather than attempting an independent deblending of the two
sources. As noted in the introduction, it is the exceptional
brightness of this target in the W3 and W4 bands that initially
motivated the follow-up observations described here.  While the
sources are blended in the W4 band, the light distribution of the host
galaxy is distorted, such that it is likely that both the GRB host
galaxy and its near neighbour are luminous at 22$\mu$m.

\begin{figure*}
\begin{center}
\includegraphics[width=1.5\columnwidth]{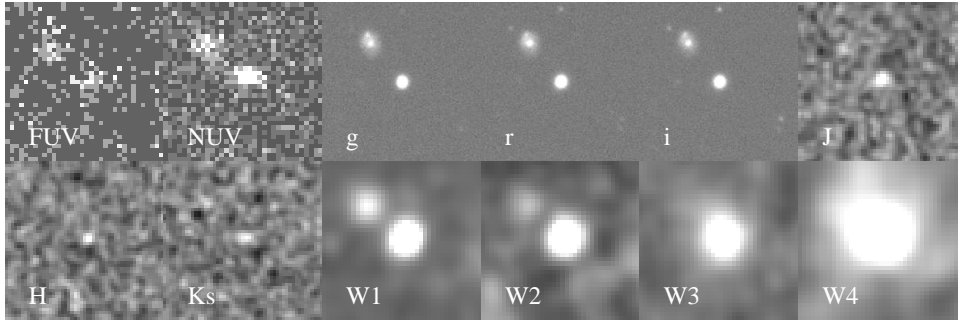}
\end{center}
\caption{Snapshot images of the host-neighbour system from the ultraviolet through to the infrared. Boxes are 50 arcseconds on a side, centred on the GRB host, and oriented with North up and East to the left. Note the exceptionally red colours of both sources, but particularly the GRB host, in the WISE (infrared) bands.\label{fig:allbands}}
\end{figure*}

The multi-wavelength photometry of both the GRB host and its neighbour
are given in table \ref{tab:phot} and figure \ref{fig:allbands} presents snapshots
of the host and neighbour in imaging across the full wavelength range. At $z=0.09$, 
the $g$-band absolute magnitude is M$_g=-20.12\pm0.05$ (comparable
  to that of the Milky Way).

\begin{table}
\begin{tabular}{lllll}
Band  &  $\lambda_\mathrm{cen}$ / \AA & Source & GRB Host  & Neighbour \\
\hline\hline
$FUV$ &    1540   & GALEX     &    20.84 $\pm$  0.20  &   20.92 $\pm$ 0.22 \\
$NUV$ &    2316   & GALEX     &    20.42 $\pm$  0.11  &   20.74 $\pm$ 0.16 \\
$g $  &    4660   & This work &    18.03 $\pm$  0.05  &   18.42 $\pm$ 0.06 \\
$V $  &    5500   & CRTS      &    17.60 $\pm$  0.09  &   18.14 $\pm$ 0.12 \\
$r $  &    6140   & This work &    17.73 $\pm$  0.02  &   18.18 $\pm$ 0.05 \\
$i $  &    7565   & This work &    17.46 $\pm$  0.01  &   18.22 $\pm$ 0.02 \\
$J $  &   12400   & 2MASS     &    17.37 $\pm$  0.13  &   17.91 $\pm$ 0.21 \\
$H $  &   16600   & 2MASS     &    17.22 $\pm$  0.15  &   19.56 $\pm$ 0.95 \\
$Ks$  &   21600   & 2MASS     &    17.50 $\pm$  0.20  &   17.91 $\pm$ 0.26 \\
$W1$  &   33500   & WISE      &    17.33 $\pm$  0.04  &   18.49 $\pm$ 0.05 \\
$W2$  &   46000   & WISE      &    17.57 $\pm$  0.05  &   19.05 $\pm$ 0.12 \\
$W3$  &  120000   & WISE      &    15.09 $\pm$  0.05  &   17.12 $\pm$ 0.32 \\
$W4$  &  220000   & WISE      &    13.68 $\pm$  0.11  &   14.78 $\pm$ 0.23 \\
\hline
Radio &   6cm     & WSRT      &   0.22$\pm$0.04\,mJy &  \\
\end{tabular}
\caption{Observed photometry measured from broadband observations of the GRB host and its neighbouring galaxy. All magnitudes are given in the AB system. Note that the neighbouring galaxy is undetected in the 2MASS $H$-band and barely detected in $Ks$. Radio fluxes are described in section \ref{sec:radio}. In the 12 and 22$\mu$m bands we make use of WISE catalog data rather than attempting to deblend the two sources independently.\label{tab:phot}}
\end{table}

We fit the spectral energy distribution (SED) of the host galaxy using a
template fitting approach, minimising the $\chi^2$ parameter to
determine the best fit age, mass, star formation history and dust
extinction.  While we could constrain the dust parameter using the
extinction derived from the hydrogen Balmer lines, we allow it to
vary, recognising that regions contributing to the continuum flux
at long wavelengths and those contibuting nebular emission flux in the
optical may
well differ in their extinction properties.

We use as templates the 
 Binary Population and
Spectral Synthesis ({\sc BPASS}) stellar population models of
\citet{2012MNRAS.419..479E,2009MNRAS.400.1019E} which include a
prescription for the nebular emission excited by the stellar
continuum. The {\sc BPASS}
models consider the instantaneous-burst and constant star formation
rate cases. In both cases, we modify the templates 
using the
\citet{2000ApJ...533..682C} dust extinction law. This was derived for
local infrared-luminous galaxies with active star formation, and would
appear appropriate in this case, given the bright infrared fluxes
measured. 

\begin{figure}
\begin{center}
\includegraphics[width=1.05\columnwidth]{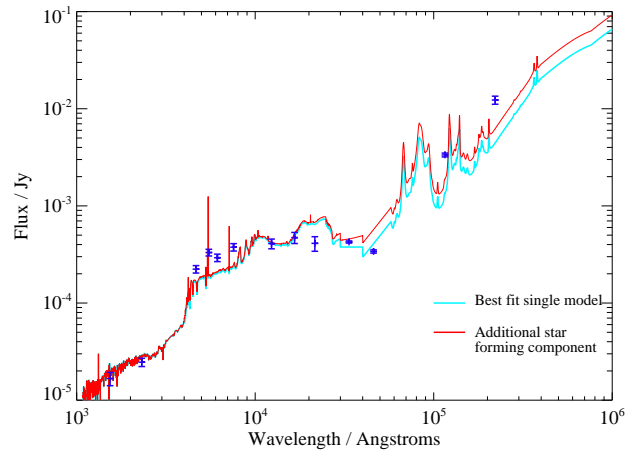}
\end{center}
\caption{The best fitting model templates to the host galaxy
  photometry. The pale cyan line indicates the best fitting single
  template: a mature system observed 0.5\,Gyr after an initial
  starburst. In red we show the best fit derived when a component of
  ongoing star formation (at age 1\,Myr, contributing just 0.1\% of
  the stellar mass) is allowed in addition to the dominant
  model.\label{fig:bpass}}
\end{figure}

The photometry shows a challenging combination of a very flat optical
spectrum (which implies little extinction and potentially even a
non-stellar continuum) and evidence for strong dust emission in the
infrared (see figure \ref{fig:bpass}). The {\sc BPASS} population
synthesis models include a treatment of stellar evolution pathways
through binary evolution. For young stellar populations, this
treatment results in a relatively blue UV-optical continuum at a given
age (and hence larger energy budget for heating dust) compared to models
which neglect such pathways.  
To model the re-emission of thermal photons at longer wavelengths,
we adopt the energy-balance prescription of
\citet{2008MNRAS.388.1595D}, and re-emit the energy lost from the
UV-optical as a combination of black body and PAH emission
components. For the latter we scale the composite mid-infrared
spectrum determined for star forming galaxies by
\citet{2007ApJ...656..770S}. The W3 and W4 bands were excluded from
the fitting procedure, in order to assess whether the observed
UV-optical continuum were able to correctly predict the mid-infrared
flux, or whether an additional, heavily dust-extincted component was
required.

The best fitting single SED model to the host photometry is not one
dominated by nebular emission or continuous star formation, but rather
a post-starburst template, observed 500\,Myr after the initial
starburst, as shown in figure \ref{fig:bpass}. The derived stellar
mass is log$_{10}$(M$_\ast$/M$_\odot$)=9.58$^{+0.12}_{-0.16}$, and a
relatively low extinction of A$_V$=0.16$\pm$0.02 is required for the
dominant stellar component.  This model reproduces the GRB host galaxy's
ultraviolet and infrared continua, while underestimating
its optical flux by $\sim25-50$\% .We note that fitting instead with the
\citep{2005MNRAS.362..799M} stellar population synthesis models
returns similar parameters, albeit with a relativity poor fit to the
data. The mass, low extinction and age of the dominant stellar
component imply that, for $z=0.09$, the GRB host represents a
relatively young, slightly sub-L$^\ast$ galaxy.

As section
\ref{sec:spec} demonstrated, there is a substantial contribution to
the optical from line emission, which may account for some part of
this discrepency. In the $r$-band in particular, line emission likely
contributes a minimum of 12\% of the broadband flux. To address this,
we also allow an additional component of continuous ongoing star
formation with moderate dust extinction (as seen in the Balmer
series), while cautioning that this may be overfitting the limited
data. We find that the star forming component required for the best
fit combination contributes a mere 0.1\% of the stellar mass of the
system. Including such a component improves the fit in the optical and
at 12 and 22$\mu$m, but causes the flux in the $K_S$ band and
4.6$\mu$m W2 band to be somewhat overestimated (again, by a factor of
$\sim$50\%). Since the latter lies at the transition between the
stellar and dust components of the template, it is possible that this
transition is not correctly addressed in the modelling, and
potentially that a steeper spectral index is required in the infrared
PAH emission region than is seen in the IR-luminous galaxy composite
used \citep{2007ApJ...656..770S}. 

Explanations for the comparatively low flux measured in the 2MASS
$K_S$ band, and the non-detection of the neighbour in the $H$-band
(see figure \ref{fig:allbands}), are less clear. It is likely that
deeper near-infrared photometry is required to address this issue.
We note that the stellar population templates fail to entirely reproduce the high fluxes seen in the
22$\mu$m W4 band, implying that at least some fraction of the stellar
emission in this system is heavily extincted and undetectable in the
UV-optical. Further observations at millimetre/submillimetre
wavelengths will also be required to properly constrain this emission
region.

\section{Star Formation in the host of GRB\,080517}\label{sec:sfr} 

The host of GRB\,080517 is an actively star forming galaxy at $z=0.09$. The evidence
for ongoing star formation is overwhelming, based on the presence of i)
strong H$\alpha$ (and other Balmer) emission lines, ii) GALEX FUV and NUV flux, iii)
22$\mu$m emission and iv) a 4.8\,GHz radio detection, not to mention the initial selection
through detection of a core-collapse gamma-ray burst.

In table \ref{tab:sfrs} we compare the star formation rates (SFRs) derived
from these different proxies. In all cases, the star formation rate
conversion used is subject to significant systematic uncertainty, but
those at 0-22$\mu$m are derived primarily for young ($<100$\,Myr)
stellar populations with continuous star formation, while that for the
radio continuum is based primarily on resolved measurements of
nearby star forming galaxies. No attempt is made to correct for dust
extinction, which may well be affecting different indicators
differently, and so these values are effectively lower limits on the
total star formation rate.

Unsurprisingly, the near-ultraviolet continuum (which is most affected
by the presence of dust extinction) gives the lowest estimate of
0.43\,$\pm$\,0.07\,M$_\odot$\,yr$^{-1}$.  In the
\citet{2000ApJ...533..682C} extinction paradigm, the continuum is
subject to 0.44 times the extinction in the nebular component, or
E($B-V$)=0.53.  This corresponds to a measured 2300\AA\ flux only 2\%
of its intrinsic value.  In fact, the emission in the GALEX band
appears to be associated with the mature stellar population (see
section \ref{sec:sed}) rather than the heavily-extincted star forming
component.

The star formation rates
derived from the optical H$\alpha$ emission line and the 22$\mu$m
continuum (measured in the W4 band) are consistent at the 1$\sigma$ level,
each estimating rates around
16\,M$_\odot$\,yr$^{-1}$.  Interestingly, this may suggest that the
1.5\,arcsecond wide slit used for spectroscopy may have captured the
majority of the flux from active star formation in the GRB host
galaxy, despite its relatively large ($\sim$2\,arcsecond full-width at
half-maximum) light distribution.

Curiously, the final star formation rate indicator, that derived from
the radio continuum at 4.5\,GHz produces a relatively low estimate at
7.6\,$\pm$\,1.4\,M$_\odot$\,yr$^{-1}$, only about half that determined
from the previous two measures, using the conversion rate determined
by \citet{2011ApJ...737...67M}, and extrapolating to 1.4\,GHz using a
radio spectral slope $\alpha=0.75$. An alternate conversion factor
\citep{2002ApJ...568...88Y} yields a similar but still lower estimate
(4\,M$_\odot$\,yr$^{-1}$).  

The synthesised beam of the WSRT at 4.8\,GHz
is insufficient to have resolved out a significant fraction of the flux in
this source, and the flux density is measured at better than 5\,$\sigma$, making
it likely that this is a genuine decrement.
Gigahertz frequency radio continuum in star forming galaxies arises
primarily from non-thermal synchrotron emission, generated by the
electrons accelerated by supernovae and their remnants. This introduces
a time delay between the onset of star formation and the establishment
of a radio continuum, the length of which will depend on the mass distribution,
metallicity and other properties of the stellar population. As a result,
the radio continuum flux density associated with ongoing star
formation rises rapidly with age of the star forming population before
stabilising at $>100$\,Myr \citep{2002A&A...392..377B}. If, then,
the young 1\,Myr starburst suggested by the BPASS models presents a
true picture of the ongoing star formation in this source, it is
possible that a strong radio continuum has yet to become established
and a radio flux - SFR conversion factor up to an order of magnitude higher might be
appropriate. Future observations at further radio/submillimeter
frequencies, and a measurement of the radio spectral slope, may help to
constrain the effect of star formation history on this estimate.

\begin{table}
\begin{tabular}{lcl}
Proxy  &    SFR/M$_\odot$\,yr$^{-1}$   & Conversion factor \\
\hline\hline
NUV continuum      & 0.43\,$\pm$\,0.07 &  Hao et al. (2011)\\
H$\alpha$ line emission & 15.5\,$\pm$\,0.4 &  Hao et al. (2011)\\
22$\mu$m continuum & 16.5\,$\pm$\,1.5 &  Lee et al (2013)\\
4.8\,GHz continuum &  7.6\,$\pm$\,1.4 &  Murphy et al. (2011)\\
\end{tabular}
\caption{Star formation rates derived from different proxies observed
  for the host of GRB\,080517. In the radio we apply the conversion
  factor derived at 1.4\,GHz, assuming a radio spectral slope of
  0.75. \label{tab:sfrs}}
\end{table}

\section{The Host and Environment of GRB\,080517}\label{sec:disc} 

The host of GRB\,080517 appears to be a compact, smooth galaxy
 in the local Universe. Given its UV-optical photometry, there would be
 little reason to expect significant ongoing star formation. Nonetheless,
  as the previous section describes, there is substantial evidence for an
  ongoing, young and fairly dusty (based on H$\alpha$/H$\beta$) 
  starburst, which likely dominates emission at $>10\mu$m.

 In this context, the presence of a near neighbour (separated by only
 $\sim$25\,kpc at $z=0.09$) is intriguing. The two galaxies have comparable
 masses (the neighbour is just $\sim$0.5 magnitudes fainter than the
 GRB host and similar in colour), and any interaction between them
 will constitute a major incident in the history of both galaxies.
 The gravitational forces caused by a near fly-by in the recent past could well
 have been sufficient to trigger the starburst detected in both
 sources - a starburst somewhat obscured by dust. It is also notable that the
 cores of both galaxies (including both components of the neighbour)
 and the GRB X-ray error circle lie along a common axis. Further
 modelling of the dynamics of this system, supported by integral field
 spectroscopy, and a search for low surface-brightness distortions in
 the morphology of both galaxies, will be necessary to confirm this
 picture, but the existing evidence is suggestive.

 Long-duration GRBs are typically associated with the peak
 light in their host galaxies, and with recent star formation
 \citep[e.g][]{2010MNRAS.tmp..479S}. However, the broadband continuum
 emission of the host of GRB\,080517 is dominated by a much older
 stellar population.  If, then, we hypothesise that the GRB is
 associated with the recent episode of star formation in this system,
 we are left with the conclusion that it occured in a dusty region
 (E($B-V$)=1.2) undergoing an intense star burst
 (SFR$\sim$16\,M$_\odot$\,yr$^{-1}$). Such dust extinction is
 consistent with the excess neutral hydrogen column inferred from the
 X-ray afterglow, and potentially with the failure of early-time
 optical observations to identify an optical transient distinguishable
 from the host galaxy.

\begin{figure}
\begin{center}
\includegraphics[width=0.99\columnwidth]{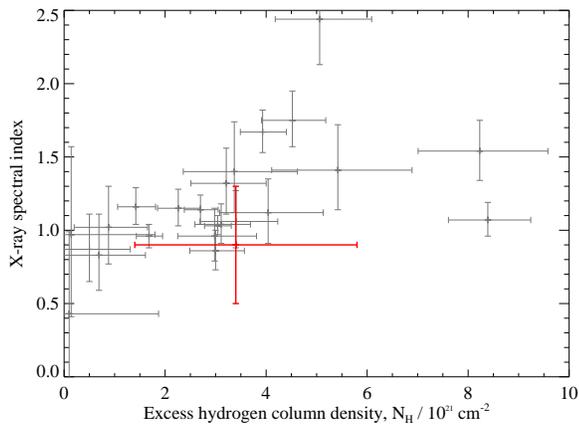}
\end{center}
\caption{The X-ray afterglow properties of GRB\,080517 compared to
  those of `dark' bursts as given by
  \citet{2013ApJ...778..128P}. GRB\,080517 (bold, red symbol) has an
  afterglow spectral index, and shows an excess neutral hydrogen
  column density above that in our own Galaxy, which is consistent
  with those of the dark GRB population, and follows the same
  correlation. \label{fig:perley1}}
\end{figure}

\begin{figure}
\begin{center}
\includegraphics[width=0.99\columnwidth]{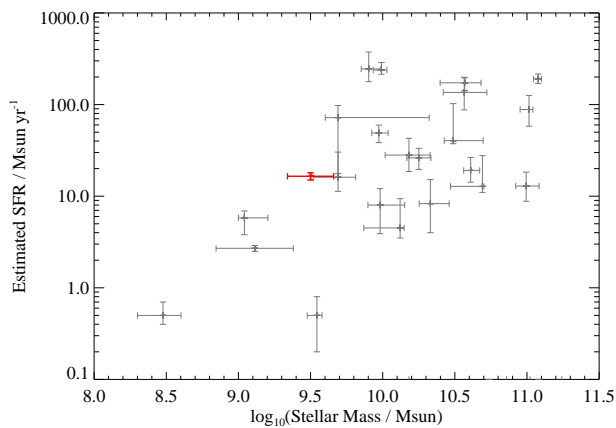}
\end{center}
\caption{The host mass of GRB\,080517, and star formation rate
  inferred from H$\alpha$ line emission, compared to those of `dark'
  bursts as given by \citet{2013ApJ...778..128P}. GRB\,080517 (bold,
  red symbol) has an inferred stellar mass (based on SED fitting)
  comparable to those of the dark burst population, and follows the
  same mass-star formation rate trend.\label{fig:perley2}}
\end{figure}

As we have already discussed, it is impossible to determine whether 
GRB\,080517 would indeed have met the `dark' burst criterion if deeper
observations were available. It is, however, possible to consider
whether its host lies in a similar region of parameter space to known
dark bursts. \citet{2013ApJ...778..128P} recently presented a systematic analysis
of the dark GRB host galaxy population, examining both their afterglow
properties and those derived from fitting of the host galaxy spectral
energy distribution.  As figures \ref{fig:perley1} and
\ref{fig:perley2} demonstrate, the inferred characteristics of the
host of GRB\,080517 lie comfortably within the distribution of `dark'
burst hosts in terms of afterglow spectral index, hydrogen column
density, stellar mass and inferred star formation rate, although the
last is higher than those in dark bursts at $z<0.5$, and more
akin to those observed at higher redshifts
\citep[][]{2013ApJ...778..128P}.

Whether or not GRB\,080517 is indeed a local example of a `dark' host, one
advantage it offers is the opportunity to study its optical spectrum in a
detail challenging for higher redshift GRB hosts of either dark or normal types.

\begin{figure}
\begin{center}
\includegraphics[width=0.99\columnwidth]{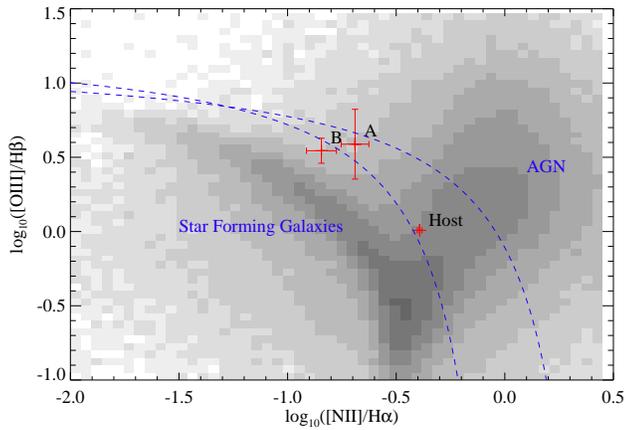}
\end{center}
\caption{The emission line strengths of the GRB host galaxy and its neighbour (components A and B) placed on the classic BPT diagram. All three sources lie above the locus of star forming galaxies measured in the SDSS, although the neighbour remains consistent with a star forming origin. The dashed lines indicate the classification criteria of \citet{2003MNRAS.346.1055K}. The region between the dashed lines is described as a `composite' region and may indicate contributions from both star formation and AGN activity. The background density plot shows the distribution of galaxies in the SDSS \citep{2004MNRAS.351.1151B}. Interestingly, the GRB host lies in the `composite' region of the parameter space, suggesting that it may have an AGN component in addition to ongoing star formation. \label{fig:bpt}}
\end{figure}

\begin{figure}
\begin{center}
\includegraphics[width=0.99\columnwidth]{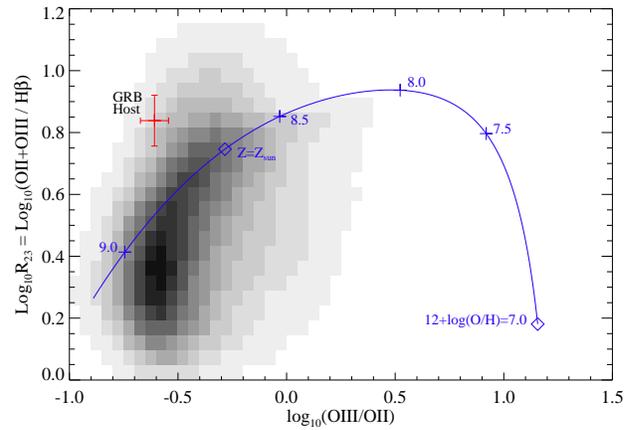}
\end{center}
\caption{Metallicity-sensitive optical line ratios for the GRB host galaxy. The well known R$_{23}$ index is plotted against the ratio of oxygen lines in order to break the degeneracy in metallicity in the former. The majority of SDSS sources (greyscale) are relatively local and high in metallicity \citep{2004MNRAS.351.1151B}. The solid line shows the metallicity parameterisation of \citet{2008A&A...488..463M}. The GRB host galaxy lies well above the typical SDSS galaxy in R$_{23}$ but within the distribution of local sources. We note that the effects of correcting for differential dust extinction on the lines is to move it further from the SDSS relation. Its measured line ratios are consistent with a slightly super-Solar metallicity.\label{fig:metal}}
\end{figure}

In figures \ref{fig:bpt} and \ref{fig:metal} we compare its optical
emission line ratios to those of  local emission-line galaxies
from the SDSS \citep{2004MNRAS.351.1151B}.  The GRB host has line
ratios consistent with a Solar or slightly super-Solar metallicity,
and is within the range of scatter of the SDSS sample, although well
above the relationship between the R$_{23}$ and [O\,{\sc
    III}]/[O\,{\sc II}] at a given metallicity determined by
\citet{2008A&A...488..463M}. This is consistent with the results from
the less metal-sensitive SED fitting procedure described in section
\ref{sec:sed}, in which 0.5-1.0 Solar metallicity templates were
narrowly preferred over those with significantly lower metal
enrichment. While far from unique
\citep[e.g.][]{2012MNRAS.420..627S,2012A&A...546A...8K},
this places the host of GRB\,080517 towards the upper end of the
metallicity distribution for GRB hosts. Interestingly, at least two other
high metallicity bursts, GRB\,020819 \citep{2010ApJ...712L..26L} and 
GRB\,080607 \citep{2009ApJ...691L..27P}, are dark bursts. 

The BPT diagram \citep{1981PASP...93....5B} is an established
indicator of starburst versus AGN character, since the different
ionisation parameters arising from the two classes have a strong
effect on optical emission line ratios and particularly the ratio of
[N\,{\sc II}] to H$\alpha$. As figure \ref{fig:bpt} demonstrates, the
two components of the neighbouring source are broadly consistent with
a star-formation driven spectrum, although both lie above the local
mean in [O\,{\sc III}]/H$\beta$ \citep[a trait also often seen in high
  redshift star forming galaxies, e.g.][]{2014ApJ...785..153M,2014arXiv1408.4122S}. While
component A has measured line ratios consistent with a `composite' spectrum,
 large errors on the measured values permit a purely star forming 
spectrum. 

The line ratios of
the GRB host galaxy are intriguing, placing it too in the region of
the parameter space usually described as `composite', suggesting that
there might plausibly be a contribution to the ionising spectrum from
an AGN.  If so, this would be a surprise, since gamma ray bursts have
not previously been associated with active galaxies, but may
 help to explain the excess flux seen in the $22\mu$m band
where a hot AGN would be expected to make a contribution to PAH
emission.

Unfortunately {\em Swift} did not track the burst beyond 20 hours
after the trigger, at which point the X-ray afterglow was still
fading. However the measured flux at this late epoch provides a firm
upper limit on possible X-ray flux from the host galaxy of
4.3$^{+2.8}_{-2.0}\times10^{-14}$\,ergs\,cm$^2$\,s$^{-1}$ in the {\em
  Swift} XRT 0.3-10\,keV band.  The luminosity of any hypothetical AGN
at $z=0.09$ is therefore constrained to an
$L_X<8.6\times10^{41}$\,ergs\,s$^{-1}$, placing it at the very low end
of the AGN luminosity distribution
\citep[e.g.][]{2008ApJ...679..118S,2010ApJ...716..348B}. As noted in section \ref{sec:afterglow},
there is also little evidence for any optical variability in the host galaxy, either before or after
the gamma ray burst, as might be expected of a galaxy with a strong AGN contibution.

While the centre of the host galaxy lies outside the 90\% confidence
interval on the X-ray location of the GRB (based on the refined XRT
analysis), the two locations are consistent at the 2$\sigma$
uncertainty level. It is therefore not impossible that the gamma-ray
burst resulted from activity in the galactic nucleus. 

A rare class of gamma ray flares are known to result from a sudden
accretion event due to the tidal disruption of stars around
supermassive black holes \citep[][Brown et al
  submitted]{2011Sci...333..203B,2012ApJ...753...77C}. The burst of
accretion in such sources launches a relativistic jet, and would
result in a short-lived burst of AGN activity from otherwise quiescent
galactic nuclei.  In this context, the plausible association of
GRB\,080517 with a very low luminosity AGN at $\sim$6 years post burst
merits further investigation and monitoring of this system.

Further observations will be required to place firmer constraints on
the presence or absence of an AGN at late times and any late time
variability.  We note that if there is no AGN contribution, then the
optical emission line ratios in the host imply star formation with a
steep ultraviolet spectrum, causing a higher ionisation parameter than
is typical at low redshifts, and perhaps strengthening the suggestion
that this is a very young, intense starburst (as suggested by the
{\sc BPASS} stellar population models).

\section{Conclusions}\label{sec:conc}

In this paper we have presented an analysis of new and archival data for the host galaxy of GRB\,080517. Our main conclusions can be summarised as follows:
\begin{enumerate}
\item  GRB\,080517 is a rare, low luminosity, long gamma ray burst.
\item  Our WHT spectroscopy reveals that the host galaxy of GRB\,080517 is a strong optical line emitter lying at $z=0.09$.
\item  The morphology of the GRB host appears to be smooth and compact with a half light radius, deconvolved with the seeing, of 2.7\,kpc. Its light distribution is consistent with a Sersic index of n=$1.5\pm1.0$.
\item  The strong optical emission line ratios in the GRB host are consistent with a composite AGN+starburst spectrum at Solar or super-Solar metallicity, and the ratio of Balmer lines suggests the nebular emission is subject to an extinction E($B-V$)=1.2.
\item  The spectral energy distribution  of the galaxy in the UV-optical is broadly reproduced by a post-starburst template at an age of 500\,Myr, with a relatively small component of ongoing star formation ($<$1\% of the stellar mass). However no template considered provides a good match to all features of the SED, and in particular to the high fluxes measued at $>10\mu$m, suggesting that multiple components with different spectral energy distributions may contribute to the broadband flux.
\item  Star formation rate estimates for the GRB host range from 0.43\,M$_\odot$\,yr$^{-1}$ to 16.5\,M$_\odot$\,yr$^{-1}$, based on different indicators. The low rate estimated from the ultraviolet continuum likely arises due to strong dust extinction in the star forming regions. Estimates from the H$\alpha$ line and $22\mu$m are consistent at 15.5\,$\pm$\,0.4 and 16.5\,$\pm$\,1.5\,M$_\odot$\,yr$^{-1}$ 
\item  We detect radio emission from the host galaxy with a flux density, $S_{4.8\,GHz}=$0.22$\pm$0.04\,mJy.  This corresponds to a star formation rate of 7.6\,$\pm$\,1.4\,M$_\odot$\,yr$^{-1}$.
\item  The high ionisation parameter seen in the optical line ratios, low radio flux and SED fitting are all consistent with a very young ($<$100\,Myr) star formation episode.
\item  The host galaxy has a close companion within 25\,kpc in projected distance and lying at the same redshift. The companion shows distorted morphology, including two cores which appear to be undergoing a merger. The proximity of these galaxies may indicate that the GRB progenitor formed in an ongoing starburst triggered by gravitational interaction.
\item  While the burst afterglow was too faint to tightly constrain the X-ray to optical flux ratio, its properties and those of its host galaxy are consistent with those of the `dark' GRB population. The host galaxy's properties and wider environment suggest that the role of galaxy-galaxy interaction in triggering bursts in relatively massive, metal rich galaxies needs to be considered more carefully. 
\end{enumerate}
 
We aim to investigate this field further, obtaining stronger X-ray
constraints on the presence of AGN activity, high resolution imaging,
and also further radio continuum measurements of the host's
dust-obscured star formation.

\section*{Acknowledgments}
We thank the anonymous referee of this paper for helpful comments and
suggestions. ERS and AJL acknowledge funding from the UK Science and Technology
Facilities Council under the Warwick Astrophysics consolidated grant
ST/L000733/1. ERS thanks Dr Elm\'e Breedt for useful discussions and
recommending the CRTS. AJvdH acknowledges the support of the European
Research Council Advanced Investigator Grant no. 247295 (PI:
R.A.M.J. Wijers). CGM acknowledges support from the Royal Society, the Wolfson Foundation and STFC.
 
Optical data were obtained from the William Herschel Telescope. The
WHT and its override programme are operated on the island of La Palma
by the Isaac Newton Group in the Spanish Observatorio del Roque de los
Muchachos of the Instituto de Astrofísica de Canarias. We also make
use of data from the Liverpool Telescope which is operated on the
island of La Palma by Liverpool John Moores University in the Spanish
Observatorio del Roque de los Muchachos of the Instituto de
Astrofisica de Canarias with financial support from the UK Science and
Technology Facilities Council.

Radio data were obtained from WSRT. The WSRT is operated by ASTRON
(Netherlands Institute for Radio Astronomy) with support from the
Netherlands foundation for Scientific Research. This work made use of
data supplied by the UK Swift Science Data Centre at the University of
Leicester \citep{2009MNRAS.397.1177E}. We also made use of Ned
Wright's very useful cosmology calculator \citep{2006PASP..118.1711W}.

Based in part on public data from GALEX GR6. The Galaxy Evolution
Explorer (GALEX) satellite is a NASA mission led by the California
Institute of Technology. This publication also makes use of data
products from the Wide-field Infrared Survey Explorer, which is a
joint project of the University of California, Los Angeles, and the
Jet Propulsion Laboratory/California Institute of Technology, funded
by the National Aeronautics and Space Administration. This publication
further makes use of data products from the Two Micron All Sky Survey,
which is a joint project of the University of Massachusetts and the
Infrared Processing and Analysis Center/California Institute of
Technology, funded by the National Aeronautics and Space
Administration and the National Science Foundation.

Data is also derived from the Catalina Real-time Transient Survey. The
CSS survey is funded by the National Aeronautics and Space
Administration under Grant No. NNG05GF22G issued through the Science
Mission Directorate Near-Earth Objects Observations Program.  The CRTS
survey is supported by the U.S.~National Science Foundation under
grants AST-0909182 and AST-1313422.

We make use of SDSS-III data. Funding for SDSS-III has been provided
by the Alfred P. Sloan Foundation, the Participating Institutions, the
National Science Foundation, and the U.S. Department of Energy Office
of Science. The SDSS-III web site is http://www.sdss3.org/.

SDSS-III is managed by the Astrophysical Research Consortium for the
Participating Institutions of the SDSS-III Collaboration including the
University of Arizona, the Brazilian Participation Group, Brookhaven
National Laboratory, Carnegie Mellon University, University of
Florida, the French Participation Group, the German Participation
Group, Harvard University, the Instituto de Astrofisica de Canarias,
the Michigan State/Notre Dame/JINA Participation Group, Johns Hopkins
University, Lawrence Berkeley National Laboratory, Max Planck
Institute for Astrophysics, Max Planck Institute for Extraterrestrial
Physics, New Mexico State University, New York University, Ohio State
University, Pennsylvania State University, University of Portsmouth,
Princeton University, the Spanish Participation Group, University of
Tokyo, University of Utah, Vanderbilt University, University of
Virginia, University of Washington, and Yale University.



\bsp

\label{lastpage}

\end{document}